\documentclass[useAMS,usenatbib]{mn2e}
\usepackage{graphicx}
\newcommand{\yy}{Y$^2$ }
\newcommand{\padova}{{\it Padova} }

\title[
Galaxy structure from the VVV DEBs -- I
]
{Tracing the structure of the Milky Way with detached eclipsing binaries
from the VVV survey --- I. The method and initial results.\thanks{Based on observations 
taken within the ESO VISTA Public Survey VVV, Programme ID 179.B-2002}} 
\author[K. G. He{\l}miniak, J. Devor, D. Minniti and P. Sybilski]
{
K. G. He{\l}miniak$^{1,2}$\thanks{E-mail:xysiek@astro.puc.cl},
J. Devor$^{3}$,
D. Minniti$^{1,4,5}$ and 
P. Sybilski$^{2}$
\\
$^{1}$Departamento de Astronom\'{i}a y Astrof\'{i}sica, Facultad de F\'{i}sica,
Pontificia Universidad Cat\'{o}lica de Chile,\\ 
Av. Vicu\~{n}a Mackenna 4860,
782-0436 Macul, Santiago, Chile
\\
$^{2}$Nicolaus Copernicus Astronomical Center, 
Department of Astrophysics, ul. Rabia\'{n}ska 8 , 
87-100 Toru\'{n}, Poland\\
$^{3}$School of Physics and Astronomy, Tel Aviv University, Tel Aviv 69978, Israel\\
$^{4}$Vatican Observatory, V-00120 Vatican City State, Italy\\
$^{5}$Departamento de Ciencias F\'{i}sicas, Universidad Andres Bello, Santiago, Chile
}
\begin{document}

\date{Accepted ..... Received ....; in original form ...}

\pagerange{\pageref{firstpage}--\pageref{lastpage}} \pubyear{2012}

\maketitle

\label{firstpage}

\begin{abstract}
We present the first results of a project aiming to trace the spatial structure 
of the Milky Way using detached eclipsing binaries (DEBs) as distance indicators. 
A sample of DEBs from the OGLE-II catalogue was selected and their near infrared
photometry was taken from the Vista Variables in the Via Lactea (VVV) survey. 
The $I$ band OGLE-II light curves are used to create models of the DEBs, which together 
with the VVV photometry are compared with a set of theoretical isochrones.
After correcting for stellar reddening, we find a set of absolute physical parameters 
of components of a given binary, including absolute magnitudes and distances.

With this approach we can calculate the distances with the precision 
better than 5 per cent. Even though we have a few systems, the distribution is not 
homogeneous along the line of sight, and appears to follow the overall structure 
of the Galaxy -- several spiral arms and the Bulge are 
distinguishable. A number of systems can be seen behind the Bulge, reaching even
the distance to the Sagittarius dwarf galaxy.
\end{abstract}


\section{Introduction}

Detached eclipsing binaries (DEBs) are extremely powerful tools 
in astronomical research. They allow us to derive a complete set of physical
parameters of stars, including their absolute magnitudes. Having them, one
can calculate the distance, or at least the distance modulus, by 
comparing the derived absolute magnitudes from the model with the observed 
ones. This however normally requires spectroscopic observations to 
calculate radial velocities of both components. They are used for 
calculating the masses and orbital parameters, crucial for obtaining
the true absolute values of the stellar parameters (like the radii).
Spectroscopic observations are more time consuming, and require much 
larger telescopes for targets of a given brightness. As pointed out 
by \citet{pac97}, one also needs accurate theoretical atmosphere models 
and/or calibrations to infer surface brightness on the basis of colours, 
line ratios or other observables. The advantage is however that the error 
in the distance estimation is comparable to, or even smaller than, the
dispersion of the $PL$-relation for Cepheids \citep{gro05}.

Nevertheless, eclipsing binaries (not only detached) have been recently used to 
determine precise distances on a different steps of the cosmic distance ladder, 
up to galaxies of the Local Group. \citet{sou05}, using various methods, 
calculated the distance to HD~23642 in the Pleiades cluster, confirming the
inconsistency between the early {\it Hipparcos} results and other findings. 
Tens of binaries were analysed in the \citep[e.g.][]{rib04,pie13} 
and SMC \citep[e.g.][]{hil05,nor10}. Finally, spectroscopy and radial 
velocities of several EBs in the Andromeda Galaxy \citep{rib05,vil10a,vil10b}
and M33 \citep{bon06} were obtained, indicating the distance modulus 
fully compatible with other findings.

Despite all this effort and the potential behind the method, eclipsing binaries 
have never been used extensively to study the structure of the Milky Way, 
especially its inner parts. Ruci{\'n}ski has been discusing the
application of W~UMa binaries for this purpose \citep{ruc97,ruc04}. 
The problem partially lays in the high and 
differential extinction towards the Galactic Center, which limits our 
magnitude range, and the crowdness of the field. Another problem is a 
lack of a deep catalogue of EBs, coming from a long-cadence variability 
survey of the Milky Way. The deepest survey so far is OGLE \citep{uda92}, 
but the only published catalogue with eclipsing binaries identified
\citep{woz02} comes from the OGLE-II phase where only selected 
fields of the Galactic Bulge were observed. EBs from this catalogue
were analysed by \citet{dev05}, but without distance determination, and
\citet{gro05} gave a list of EBs potentially useful for distance calculations,
but again without the distances themselves.

Recently, two large infra-red galactic plane surveys started: 
the UKIRT Infrared Deep Sky Survey \citep[UKIDSS;][]{law07}, with its
Galactic Plane Survey (GPS), and the VISTA Variables in the Via Lactea 
\citep[VVV;][]{min10}. The latter observes the Bulge and disk area regularly,
and the first data release is already published \citep[DR1;][]{sai12}.
The multi-epoch variability campaign in the $K_S$ band will provide
infra-red light curves for about $5\times10^5$ eclipsing binaries
in the Bulge and disk, down to magnitude $K_S\sim18$. This extensive 
and deep data set will thus be perfect to study the overall structure 
of the Galaxy, including its most inner and outer parts. The present
paper is a first effort of this study, proving that our concept works using 
a relatively small sample. We present the selected sample of
DEBs and their photometric data in Section \ref{sec_targ}. 
Sections \ref{sec_meth} and \ref{sec_res} describe the method, and our initial
results respectively. Finally in Section \ref{sec_fut}
we discuss the future possibilities and perspectives of this project.

\section{Sample and NIR data}\label{sec_targ}

\subsection{Initial target selection}
Since the main variability campaign of the VVV survey is still not finished,
we need to test our method on a well known sample of objects. We decided 
to use eclipsing binaries from the OGLE-II variable stars catalogue
\citep{woz02}. \citet{dev05} presented models of about 10000 of them
obtained with the DEBiL code, and identified 3170 of them as ``detached''. 
We selected this sample as our starting point. 
\citet{gro05} analysed the same catalogue by \citet{woz02} and
found a similar number of targets, which he describes as ``candidates
for distance estimates''. We have done all the work independently of his catalogue.

Due to some limitations of the DEBiL code (see Section \ref{sec_meth})
we decided to pre-select only the ``well'' detached systems, 
defined as the ones for which a single eclipse occupies less than about
10 per cent of the whole orbital period. We got 625 objects, for which we 
recalculated and corrected the orbital periods, using a routine
based on the analysis of variance (AoV) method \citep{s-c89}.
The routine also provides the formally second best period, which sometimes 
occurred to be the true one. We also recalculated the DEBiL models 
(with 10000 iterations, see also Section \ref{sec_meth}), especially 
correcting the total observed $I$ band magnitudes, which was originally given as 
the median value of all measurements, not the out-of-eclipse value.
As the input to the further steps we also used the 
given $V$ band magnitude (in the OGLE system), which we treat as being 
out of the eclipse. This however must be taken with caution, since 
we have no varranty that the photometry has not been contaminated 
by an observation taken during an eclipse. By selecting only the ``well''
detached systems, we expect to lower the probability of such situation
(i.e. less than 20 per cent of the systems would be contaminated).

\subsection{NIR photometry from the VVV}
The near infrared photometry comes from the merged Bulge $JHK_S$ catalogue 
of point sources, prepared as a part of the DR1 of the VVV survey 
\citep{sai12}. The catalogue (v1.1) contains over 130 million sources
from the Bulge area of the VVV survey ($-10<l<10$; $-10<b<5$),
about 70 million of which are flagged as ``stellar'' in all bands. 
The photometric measurements were done by the Cambridge Astronomical 
Survey Unit (CASU). Fluxes were measured in several apertures of different
sizes. For each target a single magnitude value is given. The catalogue
has been generated by the VISTA Science Archive \citep[VSA;][]{cro12}.

We have looked for the entries of the previously selected 625 targets 
in the merged $JHK_S$ catalogue v1.1. We assumed a rather large tolerance
in astrometry of 3 arcsec, i.e. we took only those VVV counterparts
that were not further than 3 arcsec from the OGLE position in all three
$J$, $H$ and $K_S$ bands. We found 506 matches for about 250 targets. 
Around half of the OGLE/DEBiL targets had a single match in the $JHK_S$ 
catalogue, however in the most extreme cases up to 6 matches within the 
allowed area were found. Because we wanted our procedure to be as 
automated as possible, we did not do any additional inspection or 
rejections at this stage, and decided to rely on the further steps of the
analysis and rejection criteria applied therein.

\section{The Method}\label{sec_meth}
This section describes the methods and steps we used in the analysis of the
625 DEBs, for which the initial light curve modeling was done, and later 
the 506 matches between OGLE and VVV.

\subsection{Initial light curve modeling}
The Detached Eclipsing Binary Light curve fitter \citep[DEBiL fitter;][]{dev04,dev05} 
is a fully-automated computer program\footnote{The DEBiL source code 
and running examples are available on-line at:
\texttt{http://www.cfa.harvard.edu/$\sim$jdevor/DEBiL.html}.}
which rapidly fits EB light curves to a simple EB model. Since it’s comparably 
fast and does not require any guidance from the user, the DEBiL fitter can 
systematically fit a large number of DEB light curves (LCs). 

The DEBiL fitter assumes that the binary components are well detached (i.e. that
the binary stars can be modeled as limb-darkened spheres with negligible refection), and
that 3rd light and dust reddening effects have been corrected. As input, DEBiL requires
a light curve, a period, and limb darkening coefficients; it then returns the
best fit values and estimated uncertainties for the following eight parameters:
orbital inclination ($i$), orbital eccentricity ($e$), epoch of periastron ($t_0$),
argument of periastron ($\omega$), fractional stellar radii ($r_1,r_2$), and the 
apparent stellar magnitude in the wavelength of the given LC ($mag_1,mag_2$). 
In addition to this, DEBiL provides the results of a suite of statistical tests, which
quantify the quality of the fit.

DEBiL operates by first folding the LC by the given period, then removing outliers 
and smoothing the data to reduce its noise. The shape of the resulting phased LC is then
measured to produce ``initial guess'' values for the above eight parameters,
taking into account the shape distortion due to the smoothing. Next, DEBiL
attempts to iteratively improve the model fit by running a large number of
iterations, governed by the downhill simplex optimization method \citep{nel65},
in combination with simulated annealing \citep{kir83,pre92},
for improved reliability and robustness. In the final step, the
best-solution is varied by small amounts, so to estimate the uncertainty of
each of the resulting eight model parameters.

\subsection{Binary component identification}
The DEBiL fitter attempts to fit an essentially geometric model to the LC. 
More specifically, the DEBiL parameters only describe the relative sizes and 
the orbital orientation of the binary. To better understand the physics of the 
system, one needs to also find the stellar masses, ages, and chemical composition. 
This has traditionally been done through multi-epoch high-resolution spectroscopic 
observations. Unfortunately, such observations are comparably difficult to perform 
on large numbers targets, and can be prohibitive for very dim or hot stars. 
To this end we used the Method for Eclipsing Component Identification 
\citep[MECI;][]{dev06a,dev06b}, which fits a physical model to each DEB using only
readily-available photometric data.

MECI is a fully-automated computer program\footnote{The MECI source code and 
running examples are available on-line at:
\texttt{http://www.cfa.harvard.edu/$\sim$jdevor/MECI.html}.} that is designed 
to work from the DEBiL model as a starting point, and further builds a physical 
model of the DEB. MECI assumes that the binary’s stellar components formed together 
and evolved along their respective isochrones, without any mass transfer. 
As input, MECI requires a light curve, binary colors (optional but recommended), 
a period, the orbital parameters outputted by DEBiL, and assumed initial stellar 
compositions; it then returns the best fit values and estimated uncertainties for 
binary components’ masses and coeval age. It is also possible to give specific
values of observed magnitude in different bands (with errors); MECI then returns
the absolute magnitude, so the distance may be calculated directly from the
difference between the two values.  

MECI operates by iterating through triplets of primary-component mass,
secondary-component mass, and age, and from each triplet deriving the expected stellar radii,
luminosities, and colors of the binary components. It then uses these results
to construct a model light curve and binary colors, and compares them to the
given observations. The triplet that produces the model that best matches the
observed data is assumed to be the most likely solution. Finally, the triplet
values are varied by small amounts, so to estimate the uncertainties of the
measured masses and age. Since the given isochrone library contains a finite
number of cases, it is inevitable that the best-fit triplet of some DEBs will
be at the maximal or minimal mass or age available. In such cases MECI cannot
give a bounded uncertainty estimate, but only an upper or lower limit on the
physical properties of the binary.

\subsection{Reddening-free indices and isochrones}
One of the biggest problems in calculating distances with a non-geometrical
method is the interstellar extinction and reddening. It is especially 
difficult in the direction to the galactic Bulge, not only because
of large amount of dust we are looking through, but also due to
crowding. Two stars separated by just few seconds of arc can be located 
several kiloparsecs from each other, which means completely different
values of $A_V$ or $E(B-V)$. The most popular extinction maps by \citet{sch98}
do not have enough resolution and they need to be corrected in the area of 
the Bulge. \citet{sum04} was analysing the extinction in the OGLE-II
fields, but he gave only average values of $A_V$ and $A_I$ for each field.
More recently \citet{gon11,gon12} provided a high-resolution reddening and 
metallicity map for the whole Bulge, but still the resolution is limited
to $2' \times 2'$ pixels. Finally, three-dimmensional extinction
maps were published by \citet{che13}, but they do not cover the areas where
our objects are located.

We decided to use the reddening-free indices, as in \citet{cat11}. 
They give a number of pseudo-magnitude quantities $m$,
which are combinations of magniutudes in 3 bands in the form of:
$$m_X = M_1 - c(M_2 - M_3),$$ where $M_{1,2,3}$ are the magnitudes in given bands
and $c$ is a multiplication coefficient dependent on the extinction law assumed. 
The coefficient $c$ is given in such way, that within a given extinction law,
defined as ratios of extinction values in the given bands $A_1 : A_2 : A_3$, 
the following equation is true:
$$ M_1 - c[M_2 - M_3] = (M_1+A_1) - c[(M_2+A_2) - (M_3+A_3)].$$
Moreover it can be shown that if the absolute magnitudes are used to build
the reddening-free indices, those indices
-- $M_X$ -- can be used to calculate distances 
\citep{cat11}, i.e.:
$$(m_X - M_X) = 5 (\log{d}-1),$$ 
thus they seem to be perfectly suited for the purpose of our project.

\begin{table}
\centering
\caption{Extinction law in the form of extinctions in each band 
relative to the extinction in Johnson's $V$ band -- $A_V$. }\label{tab_a}
\begin{tabular}{cccc}
\hline \hline
Filter & $\lambda_{eff}$ & $A_\lambda/A_V$ & Ref.\\
\hline 
$V_{OGLE}$& 0.542 & 1.017 & 1 \\
$I_{OGLE}$& 0.871 & 0.506 & 1 \\
$J_{VVV}$ & 1.254 & 0.280 & 2\\
$H_{VVV}$ & 1.646 & 0.184 & 2\\
$K_{S,VVV}$& 2.149 & 0.118 & 2\\
\hline
\end{tabular}
\\Ref.: (1) \citet{Van09}; (2) \citet{cat11}.
\end{table}

When building the indices of the OGLE $V,I$ and VVV $J,H,K_S$ magnitudes,
we assumed extinction ratios for each band as in Table \ref{tab_a},
which follow a canonical extinction law of $R = 3.09$,
the same as \citet{cat11}. The values of $A_\lambda/A_V$ for OGLE filters
are taken from \citet{Van09}, however they claim that those values are
reproduced by $R=2.4$. We repeated the procedure they follow \citep[from][]{car89},
and found that for given values of $\lambda_{eff}$, the extinction coefficients
$A_\lambda/A_V$ for OGLE filters are actually reproduced by $R=3.09$.
Thus our extinction law is consistent with the canonical one.

On this basis we defined five indices:
\begin{eqnarray}
m_1 &=& I - 0.563(V-K_S)\\
m_2 &=& K_S - 0.231(V-I)\\
m_3 &=& H - 1.136(J-K_S)\\
m_4 &=& J - 0.722(I-K_S)\\
m_5 &=& K_S - 1.229(J-H),
\end{eqnarray}
which, except for $m_3$, are obviously different than the ones used by \citet{cat11}.
First, we used them to produce reddening-free isochrones for MECI.
For this purpose we used two \padova sets \citep{gir00,mar08},
calculated for OGLE and VVV systems separatelly, and merged them 
into one set of a \yy format. We modified each isochrone in such a way
that instead of $V-R$, $V-I$, $V-J$, $V-H$, $V-K$ we used 
$V-m_1$...$V-m_5$ respectively.	Then, for each target we calculated
the five reddening-free indices on the basis of target's out-of-eclipse
magnitudes in the five bands (separately for each of the 506 
matchings between VVV and OGLE). 
As the input for MECI we used four ``pseudo-colours'':
$m_1-m_2$, $m_1-m_3$, $m_1-m_4$ and $m_1-m_5$, as well as the five
indices themselves (in this way only the reddening-free photometry 
was used by MECI) for calculating distance modules.
It is worth to note that in case of the reddening-free photometry 
these are the distances more reliable, not the ones given directly by 
the code in the output file. For the final value of the distance we 
took a weighted average of the five values from different indices, and
its error as a distance uncertainty.

This approach also has its disadvantages. The indices are combinations 
of three photometric measurements, each having its individual errors, which
obviously propagate and accumulate in the errors of the indices themselves. 
As it is discussed in \citet{cat11}, also the use of the indices may narrow 
the range of colours (in mag units) leading for example to more difficulties in 
recognition of certain features on colour-magnitude diagrams. 
Moreover, as mentionned before, the values of multiplication coefficitents
depend on the extinction law assumed, which is not the standard one in
the galactic Bulge. However, we still benefit from that
approach because, as shown in further sections, in the direction to the Bulge
we observe objects which are also in front of and behind it. Applying 
the values of $A_V$ or $E(B-V)$ found for the Bulge to the objects located
closer leads to further inconsistencies in the distance and absolute
magnitudes obtained \citep{rat13}. We do not know {\it a priori}
where a given system resides, as it is actually the goal of this research, 
and which extinction law should be the correct one,
so we need a more universal approach to the extinction problem, and the 
reddening-free indices seem to be the best option. 
The choice of a particular law was of course arbitrary.

The choice of the grid of metal abundances $Z$ was somewhat arbitrary as well.
The selected values are 0.0001, 0.0004, 0.0007, 0.001, 0.004, 0.007, 
0.01, 0.02, 0.03. The values below solar are similar to the ones available in the 
original \yy set, but supplemented with 0.007 and 0.0007 to better sample the
metallicity space. The only value above solar -- $Z = 0.03$ -- 
was the upper limit of the \padova set in the time the analysis 
was done. The signifficant difference between \yy and 
\padova is that the \padova models used were evolved from the ZAMS,
while \yy includes pre-main-sequence evolution. We could
obviously benefit from using isochrones with wider range of available 
ages, but the \padova set may be calculated strictly for OGLE and VVV
photometric systems, so no further transformations are required, thus the
uncertainties are reduced.

\subsection{Solution selection criteria}
For all 506 input data sets (one set for each OGLE-VVV match) 
we run MECI nine times -- one time for each set
of isochrones of a given metallicity. From the solutions of a single run
(one value of $Z$) MECI by itself selects the one with the smallest score. 
We then had to chose the value of metallicity, which provided the lowest-score
solution for a given input set. From these solutions we filtered out only those 
which met the following criteria:
\begin{itemize}
\item the MECI score, which was derived by combining the reduced chi-square 
of the best-fit model’s LC and colors, was smaller than 3;
\item all the fitted parameters were bounded (no upper or lower limits), 
and had uncertainties smaller than their values;
\item the distance uncertainties were smaller than 5 per cent.
\end{itemize}
In this way we could select those solutions which seem to be more realistic 
and secure. The 5\% relative error in distance corresponds to few hundreds of
parsecs at 5-10 kpc -- the vicinity of the galactic Bulge. This allows us to 
distinguish separate structures, like single arms. Otherwise the targets would 
be located along the line of sight more uniformly that we actually see it, 
single features would be undistinguishable, and the final results not conclusive.
In vast majority of multiple OGLE-VVV matches we managed to select most 
likely the best one, since the others gave significantly higher scores 
(worse fit). Surprisingly, it was not always the one where the VVV 
coordinates were the closest to the OGLE ones. For several cases however 
an ambiguity remained.

After that filtering we ended up with about 100 solutions. This number is small 
enough to allow for a visual inspection of the model light curves and their 
comparison with observational data. We found a number of bad fits related to 
seemingly good models, and we rejected them. We could also solve the remaining 
ambiguities, and for each target select only {\it one} reliable model, 
or reject all if none was satisfactory. 

\begin{figure*}
\includegraphics[width=0.9\textwidth]{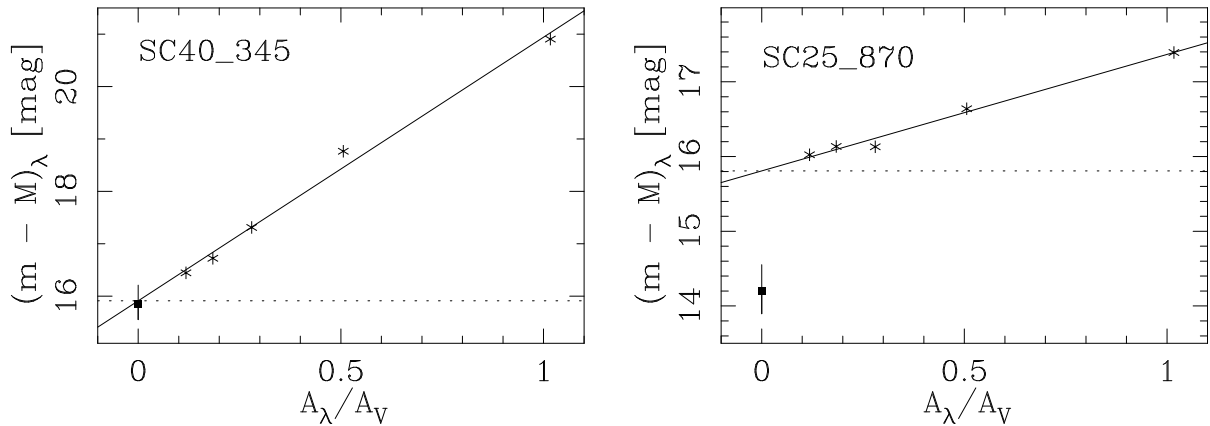}
\caption{
Examples of a solution approved (left) and rejected (right) in the last
selection step. $A_\lambda/A_V$ is the ratio of the extinction in a given band to the
extinction in Johnson's $V$, as in Table \ref{tab_a}. 
Black squares mark the distance modules related to the distances 
calculated by MECI, and their error bars refer to the 15 per cent tolerance of those distances.
Asterisks mark distance modules in bands (from left to right) $K_S$, $H$, $J$, $I$, 
and $V$ predicted by the best-fitting isochrones. Solid line represents the linear 
fit to these modules. Its slope is the extinction in Johnson's $V$ band, and its 
intercept (marked with a dotted line) is the isochrone-predicted distance modulus 
in case of no extinction. If it is consistent with the calculated one, we approve such 
a solution.
}\label{fig_avsel}
\end{figure*}

The last selection step involved checking the consistency between the
resulting distances and predictions based on the best-fitting isochrones.
For each remaining solution we generated two \padova isochrones (in OGLE and 
VVV systems) giving the best-fitting age and $Z$. Then we calculated the systems' 
total absolute magnitudes in all five bands -- $M_\lambda$, and after that 
the apparent distance modules -- $(m-M)_\lambda$. We fitted a straight line 
on the $A_\lambda / A_V$ vs. $(m - M)_\lambda$ plane, where the $A_\lambda / A_V$ 
values were taken from the Tab. \ref{tab_a}. 
In this approach, the slope of the line is simply the extinction 
coefficient in Johnson's $V$ band, and intercept with the $(m - M)$ axis is the
distance modulus without extinction $(m-M)_0$. We then translated this modulus to
a distance $(m - M)_0 = 5 (\log{d_0} - 1)$ and compared it to the distance obtained 
earlier in our fitting procedure. We rejected those solutions, for which the 
difference between two distances was larger than 15 per cent (3$\sigma$ if 
$\sigma$ is the largest error allowed).
We believe that in this way we excluded systems with large systematic
distance errors, and found reliable values of $A_V$ for the remaining ones. 
The method is presented in Fig. \ref{fig_avsel}. Approved (left) and rejected
(right) cases are shown. After this step, we finally ended up with only 23 systems,
which we present in the following Section. We list them in Table \ref{tab_coo}
together with their coordinates and photometric measurements. Their
phase-folded $I$-band light curves and models are presented in the Appendix A.

\begin{table*}
\caption{Summary data for the 23 resulting systems. DEBiL ID, coordinates
($\alpha$, $\delta$ from OGLE, and galactic $l$,$b$), VVV tile number, and OGLE-II 
and VVV photometry are given.}\label{tab_coo}
\scriptsize
\begin{tabular}{lcccccccccc}
\hline\hline
DEBiL ID & $\alpha$ & $\delta$ & $l$ & $b$ & Tile & $V_{OGLE}$ & $I_{OGLE}$ & $J_{VVV}$ & $H_{VVV}$ & $K_{S,VVV}$ \\
 & & & [$^\circ$] & [$^\circ$] & no. & [mag] & [mag] & [mag] & [mag] & [mag]\\
\hline
SC3\_2344  & 17:53:26.95 & -30:08:25.5 & -0.1580 & -2.0797 & 305 & 17.14$\pm$0.12 & 15.79$\pm$0.02 & 14.92$\pm$0.04 & 14.49$\pm$0.05 & 14.32$\pm$0.06 \\
SC10\_863  & 18:19:51.37 & -22:31:24.3 &  9.3825 & -3.5399 & 298 & 18.86$\pm$0.11 & 16.97$\pm$0.05 & 15.78$\pm$0.02 & 15.32$\pm$0.03 & 15.15$\pm$0.04 \\
SC11\_853  & 18:20:45.98 & -22:30:33.7 &  9.4936 & -3.7190 & 284 & 18.17$\pm$0.05 & 16.54$\pm$0.07 & 15.42$\pm$0.02 & 14.97$\pm$0.02 & 14.81$\pm$0.03 \\
SC11\_1200 & 18:20:54.22 & -22:21:50.4 &  9.6372 & -3.6791 & 284 & 17.49$\pm$0.11 & 16.02$\pm$0.02 & 15.51$\pm$0.02 & 15.18$\pm$0.02 & 15.07$\pm$0.04 \\
SC11\_1274 & 18:21:14.62 & -22:19:49.6 &  9.7038 & -3.7329 & 284 & 19.90$\pm$0.12 & 17.91$\pm$0.09 & 16.76$\pm$0.05 & 16.17$\pm$0.06 & 16.16$\pm$0.09 \\
SC12\_1664 & 18:16:26.54 & -24:00:08.6 &  7.7054 & -3.5442 & 297 & 19.56$\pm$0.11 & 17.41$\pm$0.19 & 16.63$\pm$0.07 & 16.28$\pm$0.10 & 16.25$\pm$0.13 \\
SC12\_3218 & 18:16:01.45 & -23:33:21.2 &  8.0541 & -3.2488 & 297 & 18.76$\pm$0.18 & 15.83$\pm$0.13 & 14.43$\pm$0.01 & 13.75$\pm$0.01 & 13.53$\pm$0.01 \\
SC15\_2498 & 17:47:44.01 & -22:59:00.2 &  5.3379 &  2.6811 & 365 & 19.32$\pm$0.13 & 17.39$\pm$0.02 & 16.19$\pm$0.07 & 15.69$\pm$0.08 & 15.54$\pm$0.09 \\
SC16\_2053 & 18:10:01.41 & -26:21:58.3 &  4.9302 & -3.3936 & 295 & 18.69$\pm$0.11 & 16.99$\pm$0.20 & 15.78$\pm$0.04 & 15.37$\pm$0.05 & 15.23$\pm$0.07 \\
SC17\_41   & 18:11:11.42 & -26:40:19.6 &  4.7866 & -3.7694 & 281 & 17.52$\pm$0.06 & 16.23$\pm$0.02 & 15.49$\pm$0.03 & 15.22$\pm$0.04 & 15.09$\pm$0.05 \\
SC18\_3886 & 18:07:16.43 & -27:03:16.8 &  4.0295 & -3.1878 & 294 & 16.27$\pm$0.03 & 15.12$\pm$0.01 & 14.20$\pm$0.02 & 13.88$\pm$0.02 & 13.80$\pm$0.02 \\
SC18\_4766 & 18:07:09.52 & -26:56:00.5 &  4.1232 & -3.1067 & 294 & 19.88$\pm$0.15 & 18.21$\pm$0.02 & 17.15$\pm$0.19 & 16.95$\pm$0.27 & 16.81$\pm$0.29 \\
SC18\_5161 & 18:06:47.19 & -26:51:15.6 &  4.1522 & -2.9957 & 294 & 17.46$\pm$0.05 & 16.13$\pm$0.02 & 15.20$\pm$0.03 & 14.85$\pm$0.04 & 14.67$\pm$0.05 \\
SC22\_2938 & 17:56:28.02 & -30:47:05.7 & -0.3876 & -2.9661 & 291 & 18.53$\pm$0.06 & 16.70$\pm$0.15 & 15.27$\pm$0.04 & 14.68$\pm$0.04 & 14.47$\pm$0.05 \\
SC22\_4501 & 17:56:29.58 & -30:31:49.9 & -0.1643 & -2.8436 & 291 & 18.15$\pm$0.06 & 15.93$\pm$0.02 & 14.46$\pm$0.02 & 13.93$\pm$0.02 & 13.67$\pm$0.02 \\
SC23\_784  & 17:57:35.48 & -31:30:55.8 & -0.9013 & -3.5393 & 291 & 17.17$\pm$0.09 & 15.47$\pm$0.01 & 14.25$\pm$0.02 & 13.86$\pm$0.02 & 13.69$\pm$0.02 \\
SC23\_1648 & 17:58:10.94 & -31:19:43.8 & -0.6762 & -3.5558 & 291 & 19.54$\pm$0.13 & 17.24$\pm$0.13 & 15.92$\pm$0.08 & 15.43$\pm$0.08 & 15.22$\pm$0.09 \\
SC27\_662  & 17:48:29.79 & -35:26:43.1 & -5.2595 & -3.8882 & 274 & 16.96$\pm$0.06 & 15.46$\pm$0.02 & 14.50$\pm$0.01 & 14.07$\pm$0.02 & 13.99$\pm$0.02 \\
SC40\_345  & 17:50:51.05 & -33:38:16.8 & -3.4546 & -3.3814 & 289 & 17.77$\pm$0.09 & 16.00$\pm$0.03 & 14.88$\pm$0.02 & 14.41$\pm$0.03 & 14.23$\pm$0.03 \\
SC41\_2400 & 17:52:23.10 & -33:01:55.7 & -2.7688 & -3.3496 & 289 & 17.81$\pm$0.15 & 15.95$\pm$0.13 & 15.19$\pm$0.03 & 14.74$\pm$0.04 & 14.58$\pm$0.04 \\
SC42\_4161 & 18:08:47.24 & -26:26:27.9 &  4.7310 & -3.1871 & 295 & 17.90$\pm$0.05 & 16.27$\pm$0.06 & 15.38$\pm$0.03 & 14.97$\pm$0.04 & 14.85$\pm$0.05 \\
SC42\_4279 & 18:09:21.47 & -26:25:51.9 &  4.8014 & -3.2941 & 295 & 18.24$\pm$0.11 & 16.32$\pm$0.04 & 15.13$\pm$0.02 & 14.68$\pm$0.03 & 14.50$\pm$0.03 \\
SC45\_1450 & 18:03:30.49 & -29:55:13.2 &  1.1177 & -3.8589 & 278 & 18.43$\pm$0.11 & 17.23$\pm$0.05 & 16.41$\pm$0.08 & 16.03$\pm$0.10 & 15.94$\pm$0.12 \\
\hline
\end{tabular}
\end{table*}

\section{Results}\label{sec_res}

\begin{figure*}
\begin{center}
\includegraphics[width=0.84\textwidth]{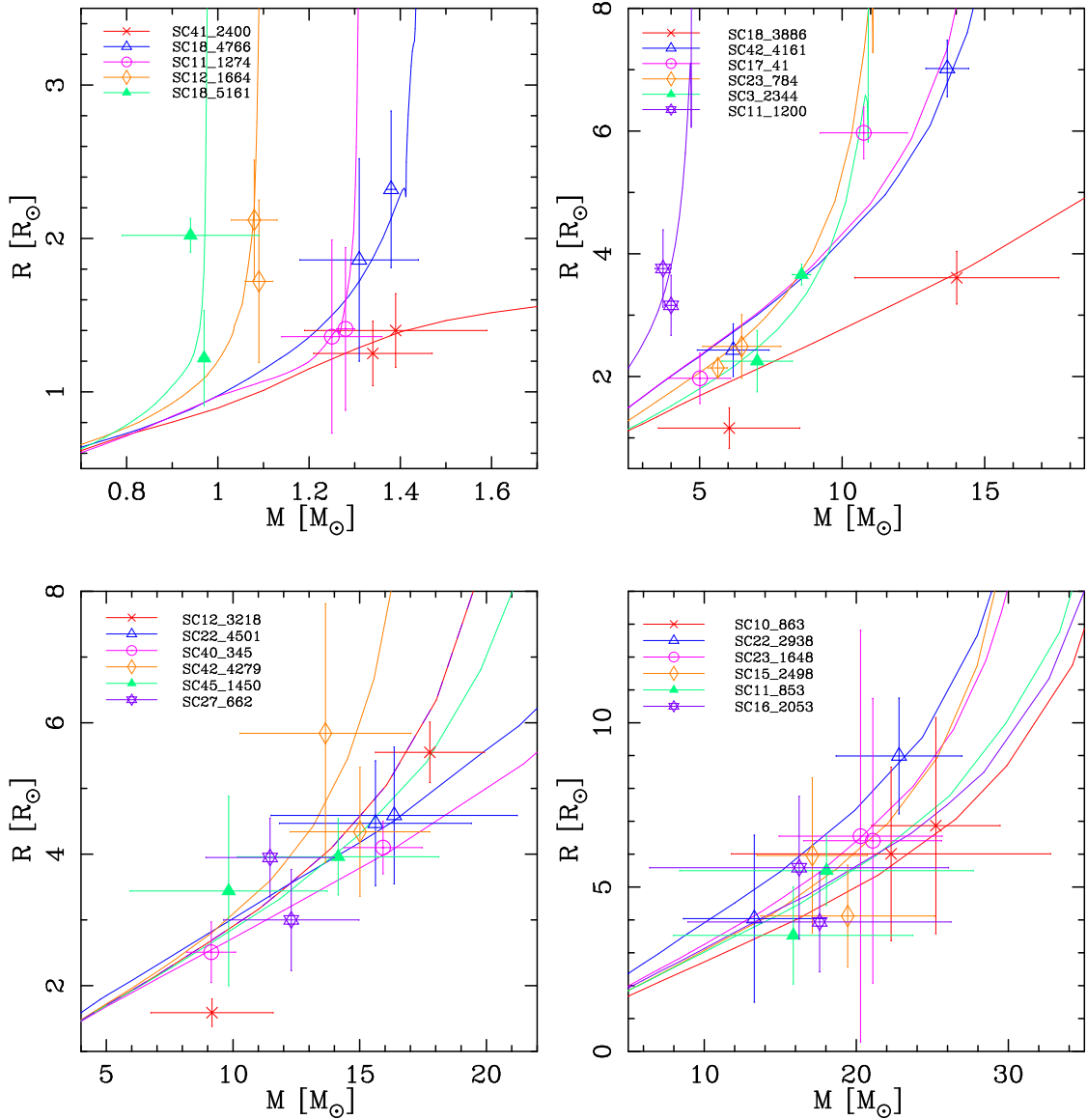}
\end{center}
\caption{Mass-radius diagram for the 23 described systems. Each system's 
components and the best-fitting isochrone are plotted in the same colour.
On the lower left pannel the isochrones for SC12\_3218 and SC27\_662
are the same. Colour figure available in the on-line version of
the manuscript.}\label{fig_mr}
\end{figure*}

\subsection{Orbital and physical parameters}\label{sec_res_par}
The output of the DEBiL and MECI codes is a set of orbital and physical 
parameters of each eclipsing system. Orbital parameters, namely the period,
absolute major semi-axis, sine of the inclination, eccentricity and
argument of the periastron (if $e>0$), are summarised in Table \ref{tab_orb}.
If the value of $e$ was smaller than its formal error (18 cases), we show it as zero, 
however it was never held fixed during any fitting. One can see that 
all the orbits may be trully circular, especially considering short 
orbital periods. For at least two of the five eccentric systems systems
the resulting eccentricity is not much bigger than its uncertainty, 
and all five have their $\omega$ indifferent from 90/270$^\circ$ -- 
a case when phases of the two eclipses differ by 0.5,
as for a circular orbit. All of the light curves, presented in
Fig. \ref{fig_lc_all}, seem to show such a situation. 

\begin{table*}
\caption{Orbital parameters of the researched systems.}\label{tab_orb}
\begin{tabular}{lcccccc}
\hline\hline
DEBiL ID & $P$ & $a$ & $\sin{i}$ & $e$ & $\omega$ \\
 & [d] & [R$_\odot$] & & & [$^\circ$] \\
\hline
SC3\_2344  & 2.526650 & 18.593$\pm$1.198 & 0.9989$\pm$0.0013 &      0.0      & --- \\
SC10\_863  & 3.471626 & 34.956$\pm$3.601 & 0.9956$\pm$0.0046 & 0.004$\pm$0.003 &  44.14$\pm$376.80 \\
SC11\_853  & 3.746524 & 32.859$\pm$5.666 & 0.9885$\pm$0.0044 &      0.0      & --- \\
SC11\_1200 & 4.445230 & 22.492$\pm$0.483 & 0.9997$\pm$0.0017 &      0.0      & --- \\
SC11\_1274 & 1.708079 &  8.198$\pm$0.143 & 1.0000$\pm$0.0047 &      0.0      & --- \\
SC12\_1664 & 2.812648 & 10.852$\pm$0.143 & 0.9960$\pm$0.0057 &      0.0      & --- \\
SC12\_3218 & 2.306802 & 22.031$\pm$1.241 & 0.9888$\pm$0.0150 & 0.140$\pm$0.017 &  90.83$\pm$94.25 \\
SC15\_2498 & 2.937350 & 28.648$\pm$2.444 & 0.9987$\pm$0.0067 & 0.040$\pm$0.010 & 275.44$\pm$104.24 \\
SC16\_2053 & 3.394874 & 30.748$\pm$5.602 & 0.9963$\pm$0.0045 &      0.0      & --- \\
SC17\_41   & 4.280280 & 27.828$\pm$1.535 & 0.9970$\pm$0.0060 &      0.0      & --- \\
SC18\_3886 & 2.427420 & 20.656$\pm$2.081 & 0.9944$\pm$0.0044 & 0.123$\pm$0.028 & 269.96$\pm$88.82 \\
SC18\_4766 & 4.873240 & 16.826$\pm$0.288 & 1.0000$\pm$0.0048 &      0.0      & --- \\
SC18\_5161 & 2.459740 &  9.520$\pm$0.255 & 0.9806$\pm$0.0052 &      0.0      & --- \\
SC22\_2938 & 4.996538 & 40.658$\pm$3.305 & 0.9957$\pm$0.0128 & 0.222$\pm$0.200  &  87.34$\pm$104.22 \\
SC22\_4501 & 2.715658 & 26.007$\pm$2.346 & 0.9845$\pm$0.0038 &      0.0      & --- \\
SC23\_784  & 3.619210 & 24.043$\pm$0.374 & 1.0000$\pm$0.0013 &      0.0      & --- \\
SC23\_1648 & 5.277104 & 44.126$\pm$3.529 & 0.9995$\pm$0.0035 &      0.0      & --- \\
SC27\_662  & 2.485568 & 22.205$\pm$1.622 & 0.9887$\pm$0.0027 &      0.0      & --- \\
SC40\_345  & 2.873292 & 24.896$\pm$0.826 & 0.9990$\pm$0.0025 &      0.0      & --- \\
SC41\_2400 & 2.361774 & 10.427$\pm$0.426 & 0.9998$\pm$0.0010 &      0.0      & --- \\
SC42\_4161 & 3.483948 & 26.199$\pm$0.882 & 0.9999$\pm$0.0138 &      0.0      & --- \\
SC42\_4279 & 3.290610 & 28.497$\pm$2.032 & 0.9874$\pm$0.0037 &      0.0      & --- \\
SC45\_1450 & 1.845630 & 18.264$\pm$1.998 & 0.9884$\pm$0.0079 &      0.0      & --- \\
\hline
\end{tabular}
\end{table*}

\begin{table*}
\caption{Stellar parameters of the researched systems.
Ages and metal abundances refer to the best-fitting isochrones.}\label{tab_CAT}
\begin{tabular}{lcccccc}
\hline\hline
DEBiL ID & $M_1$ & $M_2$ & $R_1$ & $R_2$ & $t$ & $Z$ \\
 & [M$_\odot$] & [M$_\odot$] & [R$_\odot$] & [R$_\odot$] & [Gyr] & \\
\hline
SC3\_2344  &  6.48$\pm$1.37 &  7.02$\pm$1.24 & 2.49$\pm$0.52 & 2.25$\pm$0.50 & 0.023$\pm$0.014 & 0.0001 \\
SC10\_863  & 25.23$\pm$4.22 & 22.29$\pm$10.47& 6.87$\pm$3.29 & 6.01$\pm$2.64 & 0.005$\pm$0.004 & 0.0001 \\
SC11\_853  & 18.04$\pm$9.66 & 15.85$\pm$7.87 & 5.50$\pm$1.06 & 3.53$\pm$1.48 & 0.005$\pm$0.004 & 0.0004 \\
SC11\_1200 &  3.72$\pm$0.28 &  4.00$\pm$0.22 & 3.76$\pm$0.63 & 3.16$\pm$0.49 & 0.118$\pm$0.045 & 0.0300 \\
SC11\_1274 &  1.28$\pm$0.02 &  1.25$\pm$0.11 & 1.41$\pm$0.53 & 1.36$\pm$0.63 &  2.28$\pm$1.39  & 0.0004 \\
SC12\_1664 &  1.08$\pm$0.05 &  1.09$\pm$0.03 & 2.12$\pm$0.39 & 1.72$\pm$0.53 &  7.47$\pm$3.39  & 0.0100 \\
SC12\_3218 &  9.17$\pm$2.41 & 17.77$\pm$2.15 & 1.59$^a\pm$0.21 & 5.55$^a\pm$0.46 & 0.009$\pm$0.004 & 0.0001 \\
SC15\_2498 & 17.10$\pm$3.63 & 19.43$\pm$5.72 & 5.96$\pm$2.36 & 4.12$\pm$1.55 & 0.006$\pm$0.004 & 0.0004 \\
SC16\_2053 & 16.24$\pm$9.83 & 17.58$\pm$8.65 & 5.59$\pm$2.17 & 3.94$\pm$1.52 & 0.005$\pm$0.004 & 0.0007 \\
SC17\_41   &  5.01$\pm$1.08 & 10.76$\pm$1.53 & 1.97$^a\pm$0.41 & 5.07$^a\pm$0.42 & 0.013$\pm$0.004 & 0.0040 \\
SC18\_3886 & 14.02$\pm$3.58 &  6.04$\pm$2.48 & 3.61$\pm$0.43 & 1.16$\pm$0.33 & 0.006$\pm$0.004 & 0.0001 \\
SC18\_4766 &  1.38$\pm$0.01 &  1.31$\pm$0.13 & 2.32$\pm$0.51 & 1.86$\pm$0.66 &  2.64$\pm$1.31  & 0.0070 \\
SC18\_5161 &  0.94$\pm$0.15 &  0.97$\pm$0.01 & 2.02$\pm$0.11 & 1.22$\pm$0.31 &  6.25$\pm$2.08  & 0.0001 \\
SC22\_2938 & 22.81$\pm$4.14 & 13.29$\pm$4.66 & 8.99$\pm$1.76 & 4.04$\pm$2.54 & 0.005$\pm$0.003 & 0.0070 \\
SC22\_4501 & 16.36$\pm$4.86 & 15.62$\pm$3.79 & 4.59$\pm$1.04 & 4.47$\pm$0.95 & 0.005$\pm$0.002 & 0.0004 \\
SC23\_784  &  8.58$\pm$0.32 &  5.64$\pm$0.34 & 3.66$\pm$0.17 & 2.14$\pm$0.15 & 0.023$\pm$0.003 & 0.0007 \\
SC23\_1648 & 20.28$\pm$5.36 & 21.08$\pm$4.53 & 6.55$\pm$6.26 & 6.41$\pm$4.33 & 0.006$\pm$0.002 & 0.0010 \\
SC27\_662  & 11.46$\pm$2.53 & 12.30$\pm$2.67 & 3.95$\pm$0.60 & 3.00$\pm$0.77 & 0.009$\pm$0.004 & 0.0001 \\
SC40\_345  & 15.92$\pm$1.54 &  9.14$\pm$0.96 & 4.10$\pm$0.40 & 2.51$\pm$0.46 & 0.005$\pm$0.001 & 0.0001 \\
SC41\_2400 &  1.39$\pm$0.20 &  1.34$\pm$0.13 & 1.40$\pm$0.24 & 1.25$\pm$0.21 & 0.008$\pm$0.001 & 0.0200 \\
SC42\_4161 & 13.69$\pm$0.75 &  6.18$\pm$1.26 & 7.02$\pm$0.46 & 2.43$\pm$0.43 & 0.012$\pm$0.004 & 0.0040 \\
SC42\_4279 & 13.65$\pm$3.38 & 15.01$\pm$2.75 & 5.84$\pm$1.97 & 4.34$\pm$0.98 & 0.012$\pm$0.007 & 0.0001 \\
SC45\_1450 & 14.15$\pm$3.97 &  9.83$\pm$3.89 & 3.96$\pm$0.58 & 3.44$\pm$1.44 & 0.008$\pm$0.007 & 0.0001 \\
\hline
\end{tabular}
\\Notes: $^a$ radii were inverted
\end{table*}

Other parameters, like masses, radii, ages, and metallicities 
(of the best-fitting isochrone), are given in Table \ref{tab_CAT}.
In the Figure \ref{fig_mr} we plot the 
components of every system on the mass-radius plane, together with the 
best-fitting isochrone. It is worth noting that the radii in Tab. 
\ref{tab_CAT} come from the fractional ones -- $r_1, r_2$ -- calculated 
by DEBiL, and the major semi-axis calculated with the 3rd Keplerian law on 
the basis of the known period and masses found by MECI. These are not radii
taken directly from the isochrones, so the match between the models and our 
results may serve as an independent test for the correctness of our analysis.

One can see, that the precision in determination of stellar parameters is 
not very high. The typical error is of the order of 20-30 per cent and 
reaches less than 5 per cent just for few single cases. 
Nevertheless, in most cases our parameters within errors match the 
models quite nicely, also when the two components lay far one from another
(like for SC11\_1200 or SC40\_345). In two systems -- SC17\_41 and SC12\_3218 -- 
the match was initially very bad, probably due to a well-known
issue that geometric light-curve-fitting codes, like the DEBiL, suffer 
from. In case of partially-eclipsing systems the ratio of the 
radii $r_2/r_1$ is poorly constrained, and very often degenerized with its
multiplicative inverse value $r_1/r_2$. In the same time this should not
have influenced MECI, as it uses the observed brightness of the whole 
system, which has not much to do with the fraction of the radii.
After inversing the radii, the data fit much better to the 
isochrones. Similar situation might have occurred in some other systems,
like SC27\_662 or SC42\_4279, but the errors in masses and radii are too
large to confirm that.

At this stage (with current data) we can not use our results for
detailed study of the stellar structure and evolution 
but we are able to point out potentially interesting systems which 
deserve more attention and detailed analysis, including radial velocity 
measurements and direct determination of stellar parameters.
There is for example a couple of systems with one or both components
evolving out of the main sequence. Interesting
examples are SC12\_1664 and SC18\_5161 where both components
are solar analogs, most likely beeing already sub-giants.

From Tab. \ref{tab_CAT} and Fig. \ref{fig_mr} we see that the majority
of the systems contain massive ($M > 10$~M$_\odot$) and young ($t < 10$~Myr) 
O and B type stars, sometimes after the main sequence evolution stage. 
Only few older F and G type stars can be found. In general, 
the most metal-abundant stars can be found among the least-massive 
ones, but on the other hand, the objects with the smallest $Z$ can be
found uniformly spread along the mass range. It is nevertheless clear that
high masses, young ages and metal depletion tend to be preferred in our 
solutions.

It is not very surprising in case of masses and ages. Massive, short-living
stars have high intrinsic brightnesses, so can be visible from very large 
distances. One has to remember that the OGLE-II sample is significantly limited 
in magnitude range and area on the sky, so it is expected that the further 
in distance we go, the higher fraction young and massive stars will pose.
Also the magnitude range of the VVV survey has its limitations. 
In the same time we do not expect to see many systems composed of bright,
evolved, late-type giants. They are usually found in long-period binaries, 
and among the 3170 systems classified as detached in the DEBiL catalogue, 
there are only about 60 ($<2$ per cent) systems with $P>10$~d. This is 
consistent with 0 systems in the final sample counting 23 targets. Such
late type giants will be very bright in the infrared, so many of them may
be saturated in $K_S$ and thus not included in the catalogues. 
Fig. \ref{fig_mr} shows that our objects are only main sequence stars 
and sub-giants.

The low values of $Z$ are however a bit surprising, because recent stellar 
formation is thought to occur in metal-enriched environments. The tendency of 
finding metal-poor solutions may thus be a problem of our method or the selection 
of the metallicity grid. We noticed a general weak dependency of the solution's 
score (and resulting stellar parameters) on the $Z$ of the isochrone, which for 
example makes it difficult to estimate the realistic uncertainty in $Z$, 
but in case of the metal-poorest solutions ($Z = 0.0001$) we clearly see that 
they are the formally-best ones. However, due to this weak dependency,
we do not claim that the given values of $Z$ are the true ones, rather 
that are indicating more or less significant metal depletion or enrichment 
of a given system, especially taking into account the age-metallicity 
degeneration on the main sequence. This is an issue which will be more carefully 
investigated in the future, because the recent sample of systems is too small. 
However, metal abundances in various areas of the Milky Way and surroundings 
(the membership of the systems to a certain Galaxy structures will be 
discussed in the next Section) are not well measured, or the spread is 
relatively large.

\subsection{Distances and the structure of the Galaxy}

\begin{table}
\caption{Distances, $V$-band extinctions and putative membership of 
the researched systems.}\label{tab_DIST}
\begin{tabular}{lccl}
\hline\hline
DEBiL ID & $d$ & $A_V$ & Belongs to...\\
 & [kpc] & [mag] & \\
\hline
SC3\_2344  &  8.46$\pm$0.26 & 4.029 & {\it Bulge} \\
SC10\_863  & 46.65$\pm$1.10 & 5.247 & {\it Sgr~dSph}? \\
SC11\_853  & 26.00$\pm$0.75 & 4.846 & Sag[f]/{\it Sgr~dSph}? \\
SC11\_1200 & 13.74$\pm$0.19 & 3.054 & Per[f] \\
SC11\_1274 &  6.84$\pm$0.20 & 4.127 & {\it Bulge} \\
SC12\_1664 &  9.78$\pm$0.35 & 1.461 & Far/{\it Bulge}\\
SC12\_3218 & 13.44$\pm$0.32 & 6.908 & Per[f] \\
SC15\_2498 & 40.88$\pm$1.10 & 5.320 & {\it Sgr~dSph}? \\
SC16\_2053 & 32.61$\pm$0.85 & 4.968 & {\it Sgr~dSph}? \\
SC17\_41   & 21.90$\pm$0.28 & 3.613 & ScC[f]/Sag[f]? \\
SC18\_3886 & 10.51$\pm$0.27 & 3.855 & Far/{\it Bulge}\\
SC18\_4766 & 14.16$\pm$0.59 & 2.461 & Per[f]/Nor[f] \\
SC18\_5161 &  4.29$\pm$0.04 & 2.367 & Nor[n] \\
SC22\_2938 & 30.51$\pm$1.30 & 5.618 & {\it Sgr~dSph}? \\
SC22\_4501 & 14.12$\pm$0.39 & 6.092 & Per[f]/Nor[f] \\
SC23\_784  &  7.61$\pm$0.18 & 4.804 & {\it Bulge} \\
SC23\_1648 & 43.29$\pm$0.83 & 5.904 & {\it Sgr~dSph}? \\
SC27\_662  & 12.82$\pm$0.34 & 4.421 & Per[f] \\
SC40\_345  & 14.83$\pm$0.40 & 5.029 & Nor[f] \\
SC41\_2400 &  3.21$\pm$0.02 & 2.475 & ScC[n] \\
SC42\_4161 & 27.88$\pm$0.56 & 4.381 & {\it Sgr~dSph}? \\
SC42\_4279 & 23.67$\pm$0.52 & 5.259 & Sag[f]? \\
SC45\_1450 & 33.31$\pm$0.97 & 3.851 & {\it Sgr~dSph}? \\
\hline
\end{tabular}
\\Note: 
Structures not being a part of the galactic disk are marked with italics.
The abbreviations of galactic spiral arms are as follows:
Far -- Far 3kpc Arm;
Nor -- Norma Arm;
Per -- Perseus Arm;
Sag -- Sagittarius Arm;
ScC -- Scutum-Centaurus Arm.
[f]~means the far (behind the Bulge) and [n] near part of the arm.
{\it Sgr~dSph}? means the Sagittarius dwarf galaxy; the membership is
uncertain because of actually unknown structure of this galaxy and its streams.
Question marks related to Sag[f] in 3 cases are due to a signifficant distance
from the galactic plane (1.36-1.7 kpc).
\end{table}

\begin{figure}
\begin{center}
\includegraphics[width=0.7\columnwidth]{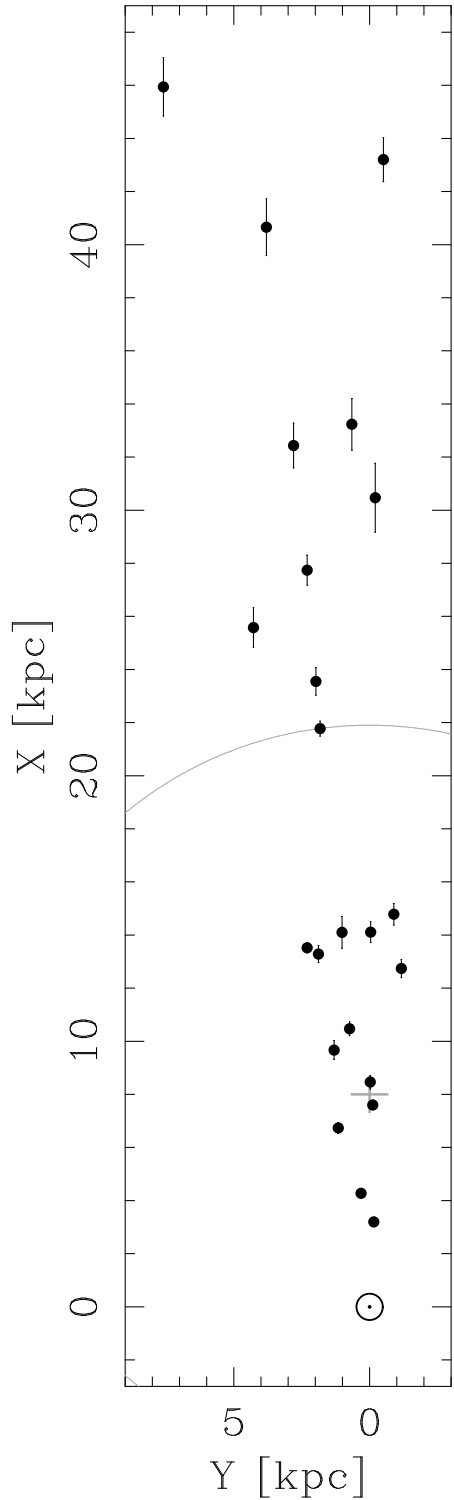}
\end{center}
\caption{Positions of the 23 researched systems in the galactic
plane. The Sun is marked at (0,0) with the traditional symbol, 
and the Galactic Center at (8.0, 0) with the grey cross. 
The grey solid line indicates an approximate edge of the Galaxy's stellar 
disk as traced by clump giants \citep{min11}.
The separation between the arms and the Bulge
is clearly seen. Most of the systems are seen behind the Bulge,
some of them may be related to the Sgr dSph galaxy.}\label{fig_xy}
\end{figure}

Calculating the distances for a sufficient number of stars and finding the 
overall structure of the Galaxy were the main goals of 
this research. We ended up with only 23 from over 3000 initial DEB's from the 
starting DEBiL catalogue, but it is enough to mark several major features of
the Milky Way and show the overall usefulness of our method. The distance,
extinction in Johnson's $V$ band, and the putative membership of a given system
to a certain structure are given in Table \ref{tab_DIST}.

The distance and the celestial coordinates can be easily transformed to 
a position in galactic coordinates $(X,Y,Z)$. In this representation the 
Sun is in (0, 0, 0), and following \citet{chu09} we assume that the 
Galactic Center is in (8.0 kpc, 0, 0). Positions of our 23 systems in 
the galactic plane $(X,Y)$ are presented in Fig. \ref{fig_xy}. 
To find their membership, we compared them to the reconstruction 
by \citet{chu09}. See also Fig. \ref{fig_rec}. 

Except for SC15\_2498, all the systems are seen around the galactic latitude 
$b\sim-3^\circ$, so the further from the Sun we go, the lower under 
the disk we look -- about 800 pc at $X\sim14$ kpc (six systems in far 
Perseus and Norma arms), and almost 3 kpc for the furthest objects. 
This is of course due to the selection of fields from the OGLE-II survey.
This obviously has its implications on studying the structure behind the 
Bulge, and may explain the low metallicities. 
Tracing the arms and thickness of the disk in this area will be 
more accurate with targets visible closer to the galactic equator.

The positions of the closest 13 binaries match very well the positions 
of spiral arms and the galactic Bulge. It is especially 
interesting in the far (behind the Bulge) part of the Perseus Arm 
($d\sim$13-14 kpc), where we put 5 systems, and for which the arm's 
curvature can be found. Three more systems are found behind the edge of 
the stellar disk found by \citet{min11} at 13.9$\pm$0.5 kpc from the Center, 
in a distance where the reconstruction by \citet{chu09} shows a weak, 
far extention of the Sagittarius arm. Their eventual
membership to this arm is however uncertain because of the large distance
from the galactic plane: from 1.36 to 1.7 kpc. 
Notice also that there are no systems found in a very strong 
Scutum-Centaurus arm behind the Bulge, about 20 kpc from the Sun. 
At the latitude of $b\sim-3^\circ$ any targets at this distance would be about
1 kpc below the galactic plane.

There is a number of systems found even futher, on distances corresponding
to the Sagittarius dwarf spheroidal galaxy (Sgr~dSph) and its stream.
If confirmed, they would be the first eclipsing binaries found in this galaxy.
Among them one can find the most massive stars of our sample. Six of the
objects classified as possible members of Sgr~dSph can be found on the lower
right panel of the Fig. \ref{fig_mr}. Two others are SC45\_1450 and
SC42\_4161. The latter, together with SC22\_2938, are the only two from
that sub-sample that have their paramater errors small enough not to
overlap on the mass-radius diagram. Yet, both have the isochrones nicely
matching the data points. SC42\_4161 is also relatively bright and its 
components are on a signifficantly different stages of evolution. 
Its spectroscopic follow up should be possible to do.

The low number of systems found in front of the Bulge -- just two -- is not 
very surprising. Due to the magnitude limitations and reduced volume sample, 
only a small fraction of all the systems contain late type stars. 
In our case it's only 5 of all 23. The two closest systems -- SC18\_5161
and SC41\_2400 -- contain F and G stars and are signifficantly brighter 
than the other three: SC11\_1274, SC12\_1664 and SC18\_4766 ($V>19.5$ mag), 
located in the Bulge and behind it. The combination of VVV and OGLE data 
thus appears to be suited much better for the poor-studied far
areas of the Galaxy than for the closer ones. 

\subsection{Ages and metallicities}

\begin{figure}
\begin{center}
\includegraphics[width=0.98\columnwidth]{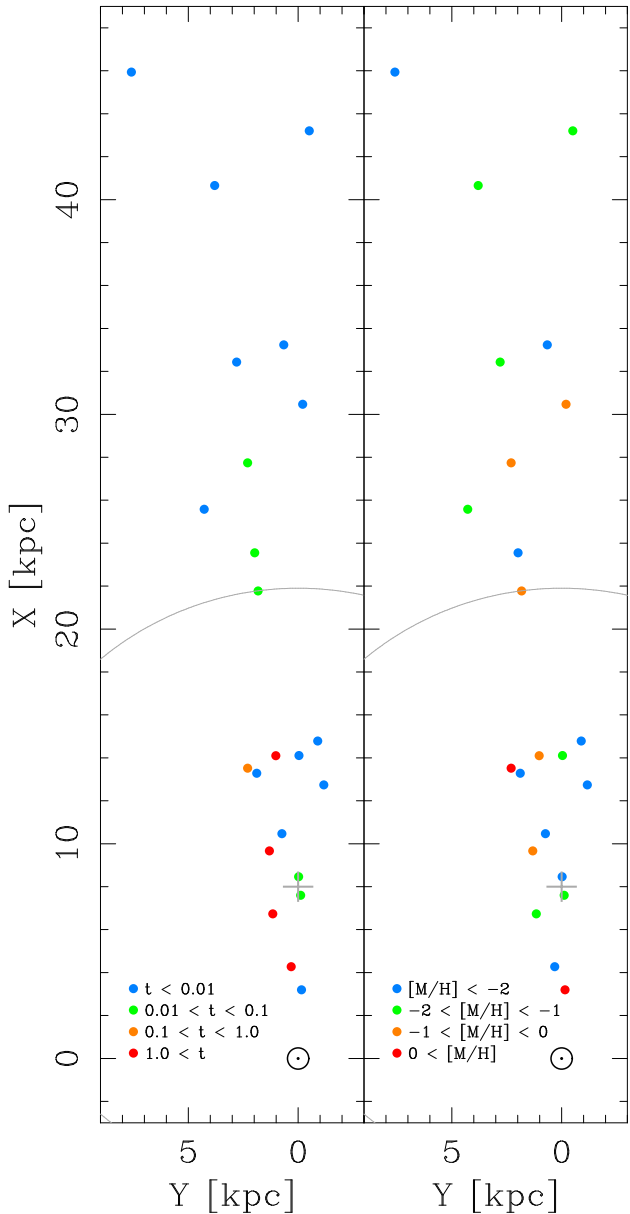}
\end{center}
\caption{Ages (left) and metallicities (right) of the reseaarched 
systems in the same spatial representation as in Fig. \ref{fig_xy}. 
The age is given in Gyr and the metal abundace $Z$ is
translated into $[M/H] = \log{(Z/0.019)}$. The Galactic Center and
the approximate edge of the stellar disk are also shown.
Colour figure is available in the on-line version of the manuscript.
}\label{fig_xy_age}
\end{figure}

Among the various output parameters of the codes, one can also find the
age and metal content of the system. As was explained preciously,
the age is calculated directly by MECI within a set of isochrones 
of a given $Z$, while the value of $Z$ was found by looking for a 
lowest score among various MECI solutions. Such an approach, at least
in principle, allows us to trace the age and metallicity distribution 
in the Milky Way. Unfortunately, the current sample of 23 is too small
to derive any conclusions, especially considering the metal-depletion
tendency mentionned in Sect. \ref{sec_res_par}.

In Figure \ref{fig_xy_age} we present the same spatial distribution
of our systems as in Fig. \ref{fig_xy} but with age and metallicity
color-coded. The age $t$ is given in Gyr and the metal abundace $Z$ is
translated into $[M/H] = \log{(Z/0.019)}$. The data does not show
any significant trends. The oldest systems are seen in arms as well 
as in the Bulge. They are not seen outside the Galaxy, but this may be
explained by the observational limitations mentionned before. 
The ages of the most massive stars are found to be of
several milions of years. This means that there 
may still exist gas and dust related to the star formation regions.
In case of the Sgr~dSph, scenarios of recent star formations
are possible, and the traces of interstellar gas have been found
\citep[e.g.][and references therein]{mon05}, but for the galactic
systems, it is surprising to see the recent
star formation so far from the galactic disk.
The metallicities are distributed more-less uniformly, regardless the
location, also within the same spiral arm. It is noticeable,
that the Sgr~dSph member candidates show quite a large spread of
metallicities -- from $[M/H] \simeq -2.3$ to $-0.4$, but as it was 
mentionned, the values of $Z$ in our case are not well constrained.

\subsection{Sources of uncertainties and incorrect solutions}

There are various factors which negatively affect our analysis, either
increasing the errors or causing that a given solution is rejected.
First one is the possibility of having a photometric measurement
in the $V,J,H$ or $K_S$ band done during an eclipse. Due to the 
pre-selection of ''well-detached'' systems only, the probablility 
of such a situation is smaller than 20 per cent, but we deal with measurements
in four filters. If even only one is affected, the resulting 
pseudo-colours and reddenig-free indices will be incorrect, so will
be the formally best solution. This may make many solutions to be 
rejected in any stage of the selection process.

Another factor, which may have the same consequences, is the inacuracy
of period determination. It is a well known issue that in cases of 
eclipses of similar depth the period-finding algorithm finds one which
is two times shorter than the true one. On the other hand, for systems
having one of the two eclipses very shallow, the period found is
twice as long as the true one. In our sample we found by eye examples of both
situations and corrected them, but obviously it is not the purpose when 
one is creating a pipeline, which should be able to work on tens of 
thousands of light curves. Moreover, the periods found may be incorrect
at the fourth or fifth decimal number. In such cases the model obtained
is less accurate than for a true period. We also found several such cases,
but we suppose that many have been omitted and analized with inaccurate 
period. We however expect that this had not as big influence on the final
results as the previous or the next factor.

The Bulge is obviously a very crowdy area, so one can expect that the
light curves will be at some level contaminated by a light of a
background star. In case of eclipsing binaries the ''third light''
changes the observed colours and makes the eclipses shallower. 
Looking at our light curves in Fig. \ref{fig_lc_all}, one can find 
cases of eclipses at the level of only 0.1-0.2 mag. While many 
''contaminated'' solutions where probably rejected due to the high 
score value or large error in distance, some of them might have 
been recognised as acceptable. Unfortunately, the fitting for 
the third light is very difficult in case of single-band light
curves, and was not implemented in DEBiL, nor in MECI.

Another possible source of uncertainties is the usage of one extinction
law for all the systems. It is already a known fact that towards the
Bulge the reddening law is non-standard and non-uniform 
\citep[][and references therein]{nat12}. Despite the 
reddening-free coefficients solve the problem of extinction and
reddening in a very elegant way, they still
depend directly on the extinction law assumed. Considering that, 
the usage of one set of coefficients to the whole analysis is an 
obvious simplification, and has an influence on the main goal of this 
study -- distance determination. Unfortunatelly, we can't rely on the
recent published extinction maps \citep{gon12,nat12,che13}, because within
one line of sight we may deal with stars located in front of, inside 
and behind the Bulge in the same time, and the reddening and extinction
should be different for each of them. To be treated correctly, the extinction
should be calculated for each target separatelly, which is not
implemented in the procedure.

As we mentionned before, the \padova evolutionary tracks we used start at the
ZAMS. This means that we will have no solutions for pre-main-sequence (PMS) 
systems, or systems containing one PMS component and the other evolving
on the main sequence already. For such objects MECI gives only upper
or lower limits of the physical parameters, and such solutions are being
rejected. It is also possible that a local score minimum have been
found in the accepted range of parameters, while the global one would be
far outside, and the value of score at the edge of the parameter space
was still higher than in the local minimum. Such a solution would still 
be accepted. This could at least partially explain the observed tendency 
to find young and metal-poor solutions, i.e. young systems far from the
galactic disk. 

Another explanation of this may be an effect of the warped 
and/or flared disk of the Milky Way, as it was used to explain the 
observed overdensities of stars in various regions of the Galaxy. 
For further reading on this topic we refer to \citet{lop12}, where a 
short summary and numerous references are presented in Section 1.
We also find possible a scenario that some of the presumably young
binaries were rejected from their parent clusters in a process
of dynamical interactions, and should have large velocities \citep{fuj12}.
This can be easily verified with spectroscopic observations.

Finally, in our analysis we also used only one limb darkening (LD) table,
prepared for the solar chemical composition \citep{kur92,cla98,cla00}. 
In general, the usage of
improper LD coefficients affects such parameters like fractional radii
or inclination. However, the precision and, what is probably more 
important, the sampling of the available light curves is usually not 
enough to be fragile for subtle changes in the LD coefficients. In
most cases we have less than 30 points per eclipse and for magnitudes
$I>17$ the spread reaches 0.1 mag. Thus we believe that the analysis of 
most of the systems was not affeected by this issue, however we can't
exclude that in case of the brightest systems.

\section{Prospects for the future}\label{sec_fut}
The final number of accepted models (23) is only a small fraction of the 
initial number of systems (3170). It is not only due to the fact that many solutions
were rejected. In the future many improvements can be done, so that we will
have hundreds or thousands of systems located uniformly in the field of the
Bulge and disk, and the structure of the Galaxy will be traced more
effectively. At this stage we mainly suffer from the inhomogeneity of the
target distribution in the sky. We rely on the OGLE-II fields, concentrated
around certain galactic latitudes. At larger distances we look below the 
disk, which makes the study of the structure of the arms difficult. 
OGLE-III covers a much larger area in the direction to the Bulge, but the 
catalogue of eclipsing binaries has not been published yet. 
This will be solved once the variability campaign of the
VVV project is finished. VVV maps uniformly the whole galactic Bulge 
($-10 < l < +10$, $-10<b<+5$) and the disk ($295 < l < 350$, $-2<b<+2$). 
The variability campaign assumes 80 epochs in $K_S$ for the whole area, 
down to magnitude 18, and $5 \times 10^5$ eclipsing binaries are expected
to be found. The photometric information also includes measurements in
two more filters than we use now -- $Z,Y$. 

The matching between the OGLE-II and VVV catalogues was relatively poor. 
Notice, that the VVV counterpart was found for less than $1/3$ of the DEBs, 
and for about half of them the matching was ambiguous. 
VVV itself should thus provide data sufficient to
effectively trace the structure of the spiral arms, especially behind the 
Bulge. On the other hand, combination of light curves in $K_S$ from the VVV
and in $I$ from the OGLE-III (once the catalog is published), and other
photometric informations, will allow for very precise study of the
overlaping regions. The structures in front, inside and behind the Bulge
(including the Sgr dSph) should then be well analysed. 

Our work should benefit not only from increasing the number of targets 
and photometric data, but also some changes in the procedures applied.
Firstly, we will extend the range of ages and chemical compositions of the isochrones
we use. This will allow us to say something not only about the distribution 
of stars, but also about ages and abundance gradients across the Galaxy.

Changes also can be made to the fitting procedures. At the current stage
DEBiL assumes spherical stars. Adding the possibility of working on 
tidaly-distorted objects will surely be very profitable. 
Only 20 per cent of the whole sample of DEBs was classified as ''well detached'',
and with $\sim$80 points per light curve in the VVV variability campaign 
we can rather expect to find 
short-period DEBs with wide eclipses and prominent elipsoidal variations.

Finally, since our approach depends on the law we assume, we expect 
that the results would change with different extinction and reddening laws,
based on different values of $R$. It is possible that the solutions we've
found are simply the ones following the law we assumed. 
The desirable solution would be to fit for extinction and reddening 
individually for each target, or implement one (probably complicated)
extinction law, suitable for any location in the Galaxy.

\section{Summary and conclusion}

We presented models and parameters of 23 detached eclipsing 
binaries (DEBs) with OGLE light curves and VVV photometric measurements.
Using the concept of reddening-free indices \citep{cat11} we
calculated distances to the systems and have been able to match
their location with major structures of the Milky Way. With the
strict criteria applied we believe to have only the most reliable
results. We conclude that our approach is suitable for tracing 
the structure of the Galaxy with DEBs identified in the VVV, 
OGLE and also UKIDS/GPS surveys. The magnitude range of 
the VVV data seem to be perfect for tracing the structure of 
spiral arms behind the Bulge, and possibly identifying the
first DEBs in the Sagittarius dwarf spheroidal galaxy and the 
associated Sagittarius stream. In principle, also the age, metallicity 
and extinction distributions can be found, but this requires many more 
objects than our initial sample contains and an uniform coverage of the
Bulge area. This may be achieved once the variability campaign 
of the VVV survey is finished.

\section*{Acknowledgments}
We would like to thank the anonymous Rereree for valuable comments,
which helped us to improve this work.

We gratefully acknowledge use of data from the ESO Public Survey
programme ID 179.B-2002 taken with the VISTA telescope, and data
products from the Cambridge Astronomical Survey Unit (CASU).
We acknowledge the support from the Basal Center for Astrophysics 
and Associated Technologies CATA PFB-06, and by the Ministry for the 
Economy, Development, and Tourism's Programa Iniciativa Cient\'{i}fica 
Milenio through grant P07-021-F, awarded to The Milky Way Millennium 
Nucleus. K.G.H. acknowledges support provided by the Proyecto FONDECYT 
Postdoctoral No. 3120153. We also acknowledge the support provided by the 
Polish National Science Center through grants 2011/03/N/ST9/01819, 
5813/B/H03/2011/40 and 2011/03/N/ST9/03192.

\appendix
\section{Light curves}

Phase-folded $I$-band light curves of the 
23 researched systems. Dots show single magnitude measurements 
and gray lines are the best-fitting models found by MECI. 
Because the solutions are done for a general case of
non-zero eccentricity, the zero-phase is set to the time of 
periastron. As it is discussed, most likely all systems have
circular orbits.

\begin{figure*}
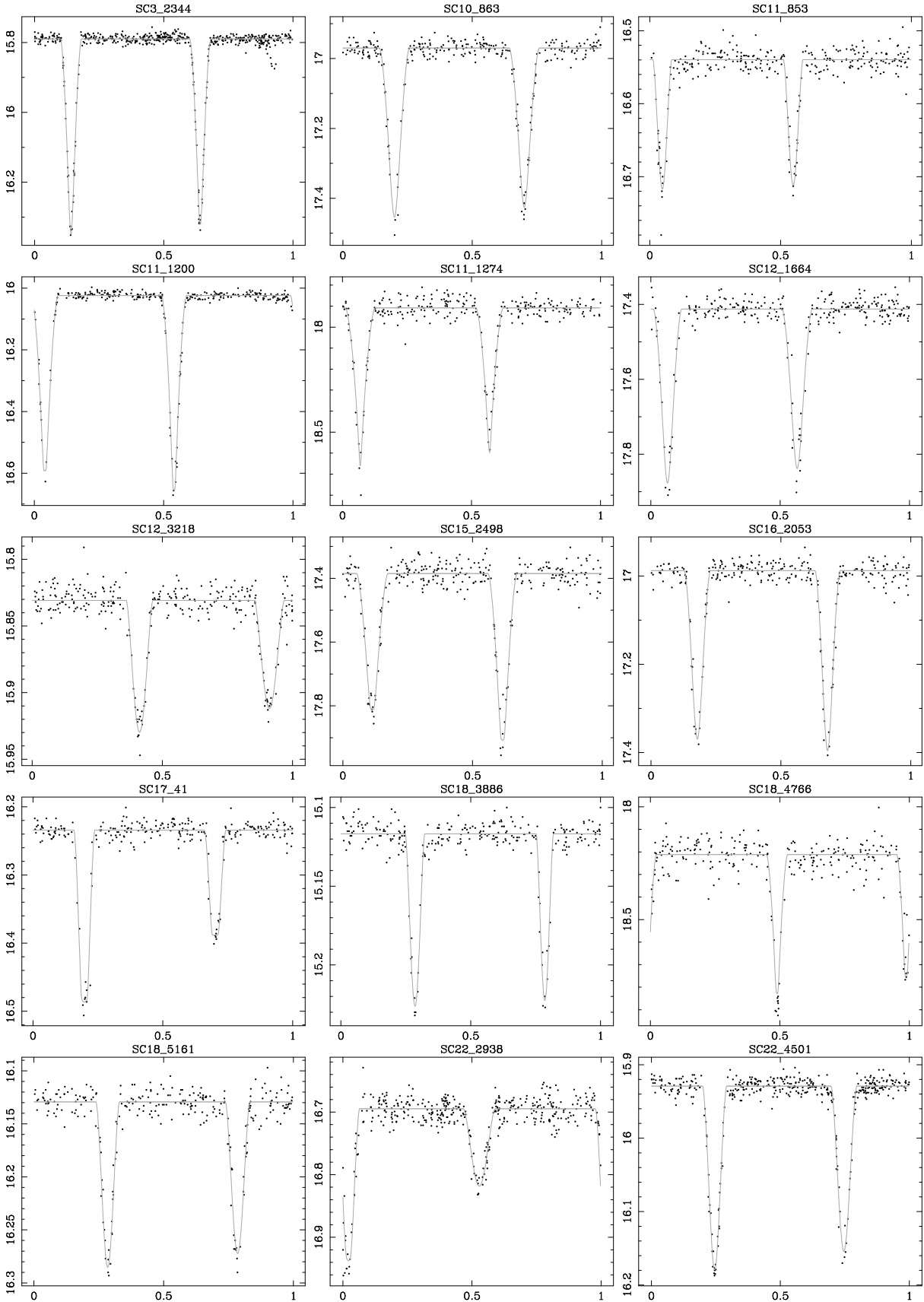

\includegraphics{lc_sc3_2344.eps}
\includegraphics{lc_sc10_863.eps}
\includegraphics{lc_sc11_853.eps}
\includegraphics{lc_sc11_1200.eps}
\includegraphics{lc_sc11_1274.eps}
\includegraphics{lc_sc12_1664.eps}
\includegraphics{lc_sc12_3218.eps}
\includegraphics{lc_sc15_2498.eps}
\includegraphics{lc_sc16_2053.eps}
\includegraphics{lc_sc17_41.eps}
\includegraphics{lc_sc18_3886.eps}
\includegraphics{lc_sc18_4766.eps}
\includegraphics{lc_sc18_5161.eps}
\includegraphics{lc_sc22_2938.eps}
\includegraphics{lc_sc22_4501.eps}
\caption{OGLE $I$-band light curves (dots) and MECI models (lines)
for the researched systems.}\label{fig_lc_all}
\end{figure*}
\begin{figure*}
\includegraphics{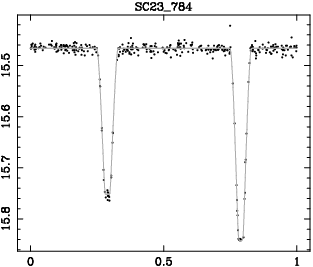}
\includegraphics{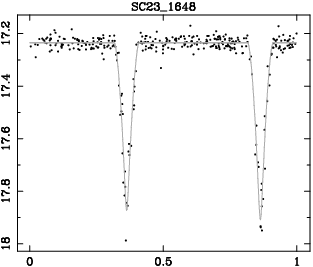}
\includegraphics{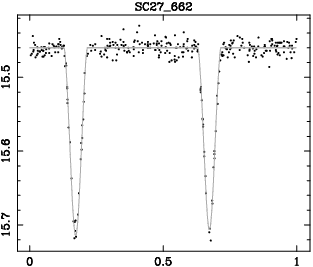}
\includegraphics{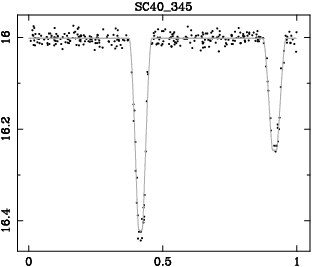}
\includegraphics{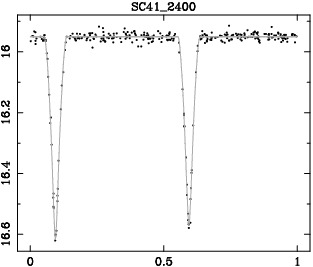}
\includegraphics{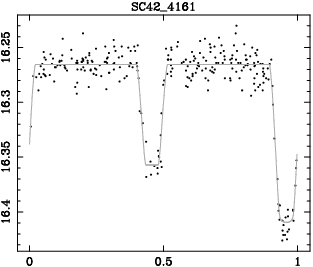}
\includegraphics{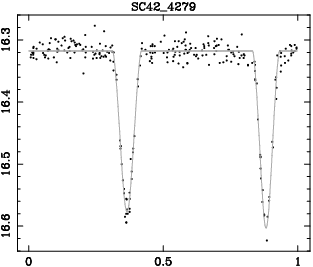}
\includegraphics{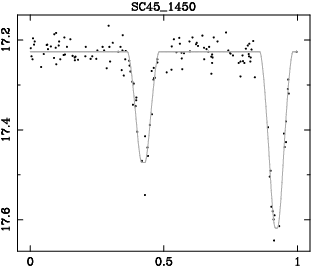}
\caption{Continuation of Figure \ref{fig_lc_all}.}
\end{figure*}

\section{Reconstruction of the Milky Way}

The 16 closest systems from our sample plotted over the reconstructed 
image of the Milky Way from \citet{chu09}. 

\begin{figure*}
\includegraphics[width=\textwidth]{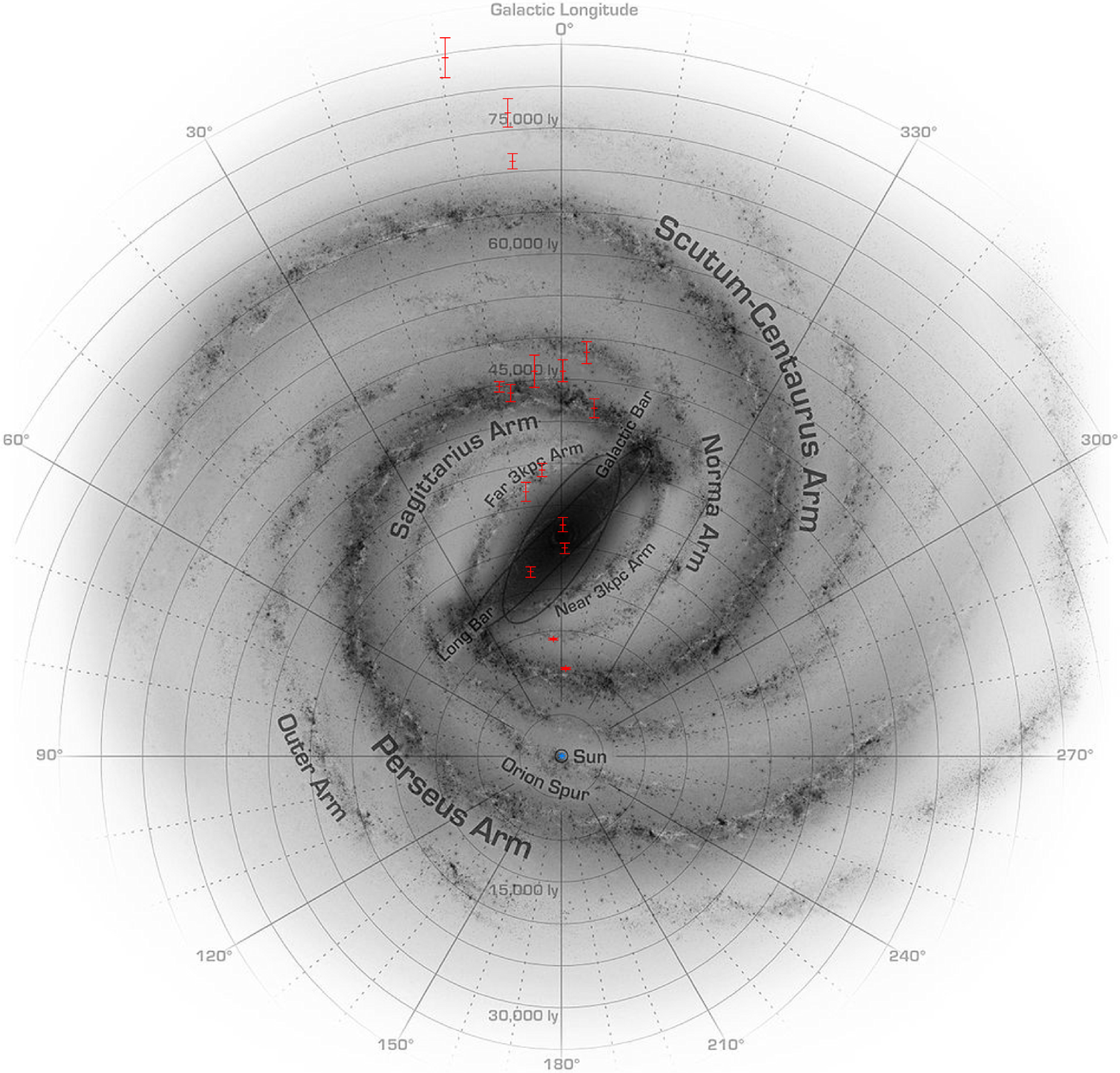}
\caption{The sixteen closest eclipsing binaries from the 23 researched plotted 
over the reconstruction of the Milky Way from \citet{chu09}. Colour figure
is available in the on-line version of the manuscript.
}\label{fig_rec}
\end{figure*}

\label{lastpage}

\end{document}